# What is the Return on Investment of Digital Engineering for Complex Systems Development? Findings from a Mixed-Methods Study on the Post-production Design Change Process of Navy Assets

Jannatul Shefa and Taylan G. Topcu

*Grado Department of Industrial and Systems Engineering, Virginia Tech, Blacksburg, VA 24061, USA*

**Correspondence:** Jannatul Shefa (shefa@vt.edu)

**Funding:** This study was supported, in part, by the U.S. Department of Navy, Navy Engineering Education Consortium (NEEC), under Contract [N00178-23-1-0003]. Any views, opinions, findings and conclusions or recommendations expressed in this material are those of the author(s) and do not necessarily reflect the views of the United States Department of Navy.

**Conflict of interest:** The authors declare they have no competing interests to disclose.

## Abstract

Complex engineered systems routinely face schedule and cost overruns, along with poor post-deployment performance. Championed by both INCOSE and the US Department of Defense (DoD), the Systems Engineering (SE) community has been increasingly looking into Digital Engineering (DE) as a potential remedy. Despite the growing advocacy, most of DE's purported benefits remain anecdotal, and its return on investment (ROI) remains poorly understood. To that end, this research presents findings from a case study on a Navy SE team conducting the preliminary design phase of post-production design change projects of Navy assets. Through a mixed-methods study, we document *why* complex system sustainment projects are routinely late, *where* and *to what extent* schedule slips arise, and *how* the DE transformation could improve schedule adherence. This study makes three major contributions. First, it documents four *archetypical inefficiency* modes driving schedule overruns and explains how these mechanisms unfold in their organizational context. Second, it quantifies the magnitude and deviation of schedule slips. Third, it creates a hypothetically digitally transformed version of the current process aligned with DoD's DE Policy, and compares it against the current state to estimate potential schedule gains. Our findings suggest that DE transformation could reduce the median project duration by 50.1% and the standard deviation by 41.5%, leading to faster, more predictable timelines. However, the observed gains are unevenly distributed across different task categories. Overall, this study presents the first quantitative evidence of DE's potential ROI and its value in improving the efficiency and predictability of complex system sustainment projects.

# 1. Introduction

The essence of systems engineering (SE) lies in the successful realization and sustainment of complex engineered artifacts that deliver stakeholder value for uncertain, and often extended, lifecycles (Blanchard & Fabrycky, 2011; Kossiakoff et al., 2011). Despite this emphasis, managing the lifecycle of large-scale complex engineered systems is often marked by recurring patterns of cost overruns and schedule delays (Government Accountability Office, 2011, 2018, 2019; Locatelli, 2018); observed across many relevant industries (Jamshidi, 2008) such as aerospace (Mavris & Pinon, 2012), defense (Oxenham, 2010), energy (Keating & Katina, 2011), and automotive (D'Ambrosio & Soremekun, 2017). The root cause often lies in the inability to manage complexity (Broniatowski & Moses, 2016; Hennig et al., 2021), the multiplicity of stakeholders and their conflicting preferences (Topcu & Mesmer, 2018), and poor decomposition decisions that fail to balance system performance, schedule, and cost (Topcu et al., 2021; Topcu & Szajnfarber, 2025).

Given this conundrum, SE research prioritizes earlier design studies (Hazelrigg, 1998); despite the fact that a significant portion of a system's benefits and costs arises during operations and sustainment (O&S) (Fricke & Schulz, 2005; Walden et al., 2015). A recent U.S. Government Accountability Office (GAO) report (U. S. GAO, 2023) documents that O&S expenses, such as platform upgrades, make up approximately 70% of a weapon system's total lifecycle cost. This is an exorbitant price tag given the scale of these portfolios. For instance, in 2020, annual O&S costs for only a subset of U.S. Department of Defense (DoD) aircraft systems topped $54B. While monetary units are generally used to document O&S shortcomings, their implications on end-user value are arguably more critical (Topcu & Mesmer, 2018). These are often manifested in capability disruptions such as increased system downtime, reduced operational availability, and mission readiness (OUSD(R&E), 2020; Sapol & Szajnfarber, 2021). For example, from 2011 to 2021, across ten Navy ship classes (e.g., frigates, destroyers, cruisers), O&S activities suffered greatly due to lack of part availability and maintenance delays; leading to a $1.2B budget overrun, along with diminished time spent in warfighter training and an overall impairment of Navy's ability to execute its mission (U. S. GAO, 2023).

To address these enduring challenges, many organizations are increasingly looking into digital transformation strategies that promise greater data integration, agility, and adaptability (Nadkarni & Prügl, 2021; Vial, 2021). In the SE community, particularly within DoD, this transformation is embodied in the Digital Engineering (DE) paradigm. DoD formally defines DE as: "an integrated approach that uses authoritative sources of system data and models as a continuum across disciplines to support lifecycle activities from concept through disposal" (Office of the Deputy Assistant Secretary of Defense for Systems Engineering, 2018). By leveraging high-fidelity authoritative sources of truth (ASOT) and integrating heterogeneous engineering models synergistically, DE aims to enable earlier detection of design flaws, facilitate multidisciplinary collaboration, and enhance the traceability of requirements and decisions (Tao et al., 2024). In DE, ASOT forms the backbone of a "digital thread" that connects every stage of the system lifecycle, enabling engineers, managers, and decision-makers to work from consistent and validated information (Department of Defense, 2023; Zimmerman et al., 2019). These capabilities are particularly critical given the long operational lifespans of complex systems and the rapid pace of technological change (Wognum et al., 2019). The aspiration is clear: to shorten development timelines, improve design quality, streamline sustainment, and reduce total ownership cost. The vision also extends beyond cost, schedule, and system performance; it aims to fundamentally change how engineering enterprises innovate, collaborate, and adapt to emergent challenges (Bone et al., 2019).

However, despite its promise, the actual return on investment of DE on improving complex system development outcomes is poorly understood (Kraft, 2020; McDermott et al., 2023). In the absence of this knowledge, many engineering organizations struggle to justify large-scale DE investments beyond anecdotal success stories (Barykin et al., 2024). In particular, there is a limited understanding of *how*, *where*, and *by how much* DE can tangibly improve SE processes. These questions are particularly relevant for sustainment and post-production modification efforts, where new technologies are integrated into legacy systems within constrained timelines and operational environments. To that end, this research investigates how schedule delays materialize in current post-production design change processes and then examines how the adoption of DE could help overcome these challenges. More specifically, this study addresses the following research questions (RQs):

**RQ1:** *Why* do post-production design change projects deviate from their planned schedule?

**RQ2:** *Where* in the process do schedule slips occur, and *how* substantial are these deviations?

**RQ3:** If the organization were digitally transformed as envisioned, *where*, *how*, and *to what extent* would the return on investment of DE be realized on project schedule?

To answer these questions, we adopted a mixed methods approach. First, we conducted a case study focusing on the preliminary design phase of post-production design change processes of an Integrated Design Team (IDT) at an anonymized U.S. Navy Agency. We collected both qualitative and quantitative data regarding the nature of the process. We then organized our data and conducted a qualitative analysis of the process sequence, which revealed recurring *inefficiency archetypes* that could be generalizable to other complex system sustainment efforts. These include delays from manual information exchange, incomplete or misaligned information and requirements, lack of up-to-date configuration data, and absence of lifecycle operation models. We then supplemented these insights with a two-fold quantitative effort. First, we adopted an *ex-post* approach and structured a simulation of the process that categorizes different SE activities and executed a Monte Carlo analysis to study the source and magnitude of delays in the current state of the process. Second, we adopted an *ex-ante* approach and constructed a hypothetical digitally transformed version of the process that is in line with DoD DE policy goals. We then compared this digitally transformed version of the process against its current structure to quantify where and by how much DE transformation could help alleviate inefficiencies.

Our findings suggest that DE transformation could reduce median project duration by 50.1%, along with a significant reduction in the predictability of the process. However, these gains are unevenly distributed across various SE activities, ranging from reductions of ~89% to ~14% in median duration in the particular process we studied; indicating that some SE activities may greatly benefit from DE transformation while others might experience relatively lower yet still substantial gains. Nevertheless, collectively, our findings bridge the knowledge gap regarding why schedule slips occur more frequently than expected in complex system sustainment and how DE transformation could help. We contend that these insights will enable practitioners, managers, and policy makers to better understand the value of DE and help prioritize their objectives.

The remainder of this paper is structured as follows. Section 2 reviews the relevant literature on SE challenges, sustainment, and the scope and benefits of DE. Section 3 describes the mixed-methods research approach, including the case study context, data collection, qualitative, and quantitative analysis methods employed. Section 4 presents the findings, integrating qualitative insights and simulation-based evidence. Section 5 discusses the findings in relation to the research questions and outlines key contributions, limitations, and practical implications. Finally, Section 6 concludes the paper, summarizing the key insights and directions for future research.

## 2. Literature Review
### 2.1 SE Challenges and Lack of Attention on Sustainment

A foundational principle of SE is to focus not only on the successful design and realization of complex systems but also on their long-term sustainment, ensuring that they continue to deliver stakeholder value over their lifecycles (Blanchard et al., 1990; Kossiakoff et al., 2011). In the SE literature, O&S is a particularly challenging problem compared to earlier design studies, as once deployed, many complex engineered systems are operated for longer than initially planned operational lives (Blanchard, 2004; Zhang & Zhang, 2014). This introduces deviations from the planned operational context for which the system has been originally optimized for, in addition to the post-production design changes large systems of systems typically need to go through to ensure mission fitness (Walden et al., 2015). These postproduction design changes are also necessary to address obsolescence, operational, and maintenance issues to accommodate evolving mission requirements, and facilitate new capabilities (Ross et al., 2008; Walden et al., 2015).

The O&S challenges, especially when design changes are required, are multifaceted (Fricke & Schulz, 2005) and are inherently sociotechnical. For instance, successful sustainment demands coordination among multiple associated stakeholders, including system developers, end-users, contractors, and mission agencies, among others. Managing the coordination among these stakeholders inherently involves crossing organizational boundaries, which in turn is documented to introduce schedule delays (Brainard & Szajnfarber, 2019). Moreover, engineered systems are often operated for extremely lengthy operational lives that may exceed 70 years in some cases, such as the B-52. As a result, the developers of the system, e.g., the subject matter experts, SEs, and program managers, retire or leave the workforce in the meantime (Cappelli, 2014; Topcu & Szajnfarber, 2023). Furthermore, the employed engineering tools and methods also become outdated (Mellal, 2020), leaving gaps in institutional knowledge and capabilities. This is also true for the modality of stored data (e.g., hand-made engineering drawings of a Cold War era ballistic missile still in operation), as both digitization and digitalization remain a huge challenge for many organizations (Tao et al., 2024; Thun et al., 2022). Finally, in many organizations, configuration control is not perfect, leading to inconsistencies along with the need for various design artifacts of the same system of interest, such as its CAD files and model-based systems engineering (MBSE) models.

Particularly in the case of complex systems-of-systems, such as aircraft carriers, aforementioned sustainment and design challenges are further complicated by the need to integrate a wide range of systems with competing performance and functional requirements (Maier, 1998; Sage & Cuppan, 2001; Raz et al., 2017). These assets also face intricate operational interdependencies, including human-systems integration, electromagnetic interference, flight operations, and other environmental constraints (Törngren & Grogan, 2018). Managing such complexity demands the involvement of large, cross-disciplinary teams that can collectively address the uncertainties and tightly coupled trade-offs inherent in these efforts (De Weck et al., 2007). To summarize, the system sustainment challenge lies at the core of the SE puzzle and requires a sociotechnical perspective that looks across the organizational and technical architectures.

## 2.2 Digital Engineering: Scope and Benefits

DE has emerged as a model-centric and data-driven approach that seeks to overcome the limitations of traditional document-based systems engineering SE (Tao et al., 2024). Rather than relying on static documents and fragmented models, DE seeks to establish authoritative digital representations as centralized knowledge repositories, ensuring explicit linkages between requirements, design changes, and verification activities (Blackburn et al., 2022; Bone et al., 2019). This integrated approach reduces information silos and provides a consistent, validated source of truth across the entire lifecycle (Huang et al., 2020a). A key element of this approach is the digital thread, which enables seamless connectivity across lifecycle stages, from concept to disposal, through synchronized models spanning CAD, multiphysics simulations, software, and SysML representations (Dunbar et al., 2023; Office of the Deputy Assistant Secretary of Defense for Systems Engineering, 2018). Historically, these artifacts were created in isolation using incompatible standards, limiting interoperability and knowledge reuse (Hedberg et al., 2019). DE addresses these barriers by establishing a single, authoritative source of system data (Zimmerman et al., 2019). More specifically, the DoD formalizes this shift through five DE policy goals: *"(1) formalize the development, integration, and use of models to inform enterprise and program decision making; (2) provide an enduring ASOT; (3) incorporate technological innovation to improve the engineering practice; (4) establish a supporting infrastructure and environments to perform activities, collaborate and communicate across stakeholders; and (5) transform the culture and workforce to adopt and support DE across the lifecycle* (Office of the Deputy Assistant Secretary of Defense for Systems Engineering, 2018)."

Most reported benefits of DE relate to agility and efficiency. By enabling rapid design iterations, engineers can explore alternatives, perform trade-off studies, and evaluate performance across multiple scenarios early in the lifecycle, reducing reliance on costly physical prototypes (McDermott, 2021). DE also supports collaborative engineering, allowing geographically distributed teams to work concurrently on synchronized models, an essential capability for system-of-systems projects (Bone et al., 2019; Farshad & Fortin, 2023). Nevertheless, the advantages of DE lie in its ability to extend beyond system design into sustainment and lifecycle management. DE introduces a closed-loop feedback mechanism between operational data and design decisions through technologies such as digital twins (Bickford et al., 2020; Bing et al., 2019; Di Maio et al., 2018), allowing systems to evolve dynamically in response to real-world conditions rather than static assumptions (Kasper et al., 2024). Furthermore, as systems grow more complex and interconnected, DE is increasingly seen as a foundational enabler for addressing sustainability challenges, resilience, and adaptability in system-of-systems contexts (Caldera et al., 2022). These capabilities make DE particularly relevant in contexts characterized by complexity, uncertainty, and frequent post-deployment modifications.

Government and industry bodies, including the DoD and International Council of Systems Engineering (INCOSE), have positioned DE as a strategic enabler of cost control, adaptability, and mission readiness, citing benefits such as reduced schedule delays, improved reliability, and enhanced responsiveness to change (Department of Defense, 2023; INCOSE, 2023; Office of the Deputy Assistant Secretary of Defense for Systems Engineering, 2018). In effect, DE represents an Industry 4.0-driven evolution of SE practices, replacing document-heavy workflows with an integrated digital ecosystem that strengthens lifecycle continuity, transparency, and compliance (McDermott, 2021; Possehl et al., 2022). However, these advantages often come with significant organizational challenges (Zimmerman et al., 2019). Implementing DE typically requires substantial cultural transformation, which in many cases is driven by top-down mandates rather than grassroots change, raising questions about sustainability and adoption (Assef & Geiger, 2023; McDermott, 2021).

Beyond technology integration, DE adoption introduces workforce and cultural challenges (Office of the Deputy Assistant Secretary of Defense for Systems Engineering, 2018). Successful implementation depends on proficiency in model-based methods, systems modeling languages, and data analytics, as well as the ability to collaborate across disciplines (Bone et al., 2019). These evolving skill requirements position DE as both a technical and organizational transformation, necessitating new strategies for training, leadership engagement, and knowledge management (Oludapo et al., 2024). Ultimately, technological readiness alone is insufficient without corresponding organizational adaptability. To summarize, DE transformation is a socio-technical challenge that necessitates a holistic approach for success (Xames & Topcu, 2025).

## 2.3. The Research Gap

Despite the widespread advocacy for DE, its direct impact on SE outcomes, particularly in terms of program schedule, remains poorly quantified. In other words, similar to the earlier experiences with MBSE, enthusiasm for DE often runs ahead of empirical evidence (Henderson & Salado, 2021). Current evidence on the value of DE is largely anecdotal, leaving open questions about where and to what extent DE delivers measurable, evidence-based improvements (Barykin et al., 2024; Kraft, 2020; McDermott et al., 2023). Although recent initiatives have proposed metrics, such as defect density, rework rates, ease of implementing changes, system understanding, and collaboration effectiveness (Henderson et al., 2023), these remain largely theoretical without widespread adoption and validation. Recent studies underscore the challenges of assessing DE's cost-benefit, noting that without consistent definitions and data tracking, such analyses remain speculative (Clayton et al., 2024; Hedberg et al., 2019; Whitehead et al., 2024).

Additionally, although the O&S phase of the system lifecycle accounts for the largest share of incurred costs while directly shaping mission readiness and success, it remains relatively underexplored within the SE literature (Fricke & Schulz, 2005; Ross et al., 2008; Sapol & Szajnfarber, 2021; Topcu & Shefa, 2025). These gaps are especially pronounced in post-production design efforts and the sustainment of legacy system-of-systems, where design changes are subject to strict constraints regarding schedule, cost, and availability. To address this gap, this study investigates current inefficiencies in post-production design change workflows and examines where time losses emerge, the extent of these inefficiencies, and how DE adoption could generate tangible benefits.

## 3. Methodology

In this study, we adopted a mixed-methods approach that synergistically combines a qualitative case study with a two-fold quantitative simulation-based analysis. Figure 1 provides an overview of the employed methodology, and the following subsections are dedicated to each of the steps, respectively. To elaborate, Section 3.1 describes the research setting, including the context and the characteristics of the examined case, and discusses data collection and its processing for this research. Section 3.2 presents the qualitative analysis employed to extract the existing process structure and document the inefficiency archetypes. Section 3.3 describes the *ex-post* simulation-based approach that leverages the current process structure to study magnitude and variations in delays. Finally, section 3.4. describes the second layer of simulation, which presents an *ex-ante* analysis that compares the current process against a future state of its post-DE transformation efforts. This section also describes the assumptions that enable such analysis.

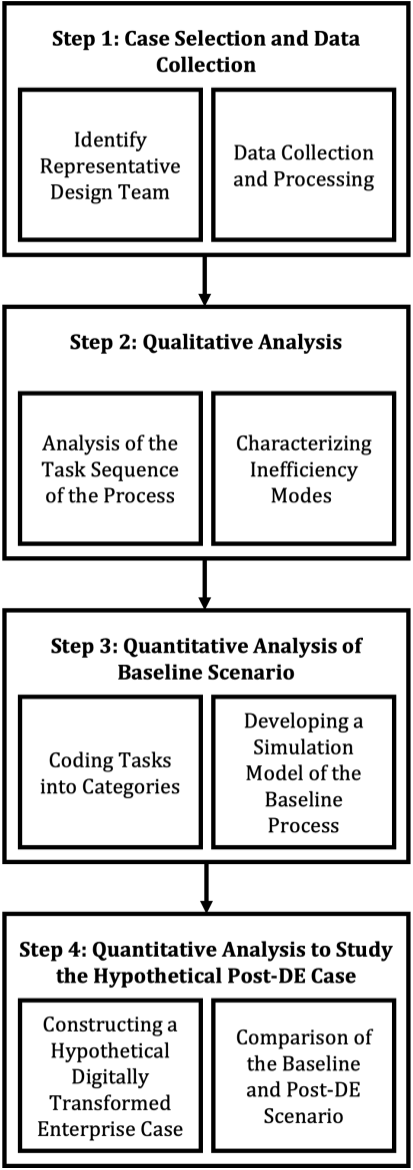

**Figure 1.** Overview of the Research Methodology.

## 3.1. Case Selection, Data Collection, and Processing

This article is based on a case study of the preliminary design phase of the post-production design change processes carried out on Navy Assets. Case studies offer value by enabling an in-depth exploration of complex phenomena in their unique context, particularly when behaviors or activities being studied cannot be manipulated. This allows the pursuit of *how* and *why* questions as processes naturally occur and is ideal for capturing the nuanced interactions and dynamics within real-world contexts, such as complex engineering systems. Additionally, case studies are effective for building or testing theories from evidence, especially in fields where existing literature and empirical evidence are limited (Eisenhardt, 1989, 2021). In this study, the methodological choice of a case study was particularly fitting because the host organization was receptive to sharing information as they were executing their daily activities, which allowed for a detailed examination of the general characteristics of the process being examined. Additionally, the research setting is representative of other early-stage design activities that DE transformation aims to support as it provides a SE team that: (i) is responsible for executing design changes in complex systems of systems operating for an extended period of time, (ii) interacts with multiple stakeholders with often conflicting interests, and (iii) makes interdisciplinary decisions utilizing diverse analytical models.

More specifically, we study an anonymized IDT and its interactions with the broader Navy engineering ecosystem. IDT is tasked with both designing, modifying, and sustaining (through post-production design change projects) the topside arrangement (i.e., above the waterline) of Navy assets. This includes a broad range of Navy artifacts that range from unmanned vessels to aircraft carriers. The team is cross-disciplinary and is composed of experts in electromagnetics, mechanical, structural, and SE, along with some retired Navy veterans who provide operational knowledge, among others (Vincenti, 1992; McGowan et al., 2012). From an organizational architecture perspective, IDT's role in the broader Navy support system is similar to a "bus node" in engineered systems. They lead design projects by engaging and leveraging the expertise of other technical and operational specialties distributed across Navy agencies.

Consequently, the IDT decisions are inherently complex and interdisciplinary. These decisions involve the arrangement of limited topside design space while facilitating counterbalancing mission objectives. All the on-board systems that need to be placed topside, such as sensors, weapons, command & control systems, and manned environments, along with their functional and performance needs, create a highly interdependent puzzle with many constraints, along with emergent properties that need to be carefully balanced. These are often hard to predict or know precisely *a priori*. Thus, IDT processes exhibit many iterations that involve the acquisition of information from both the Force and other technical stakeholders (e.g., shipyards, government agencies, contractors).

Additionally, although IDT plays a critical role in ensuring resulting designs will support the Navy's objectives, its "zone of control" is constrained by organizational boundaries, which in turn influence the efficacy of its decisions. For instance, below-deck decisions that bound the topside design space by determining engine power and deck size are delegated to another Navy organization. While this is understandable given the scale of these tasks, it effectively serves as a design constraint, limiting the design space of IDT. Another constraint is their ability to transparently monitor system lifecycle costs. As is, IDT does not exercise any cost calculation; these responsibilities are delegated to other Navy agencies. These organizational settings shape IDT processes and influence the efficacy of individual program outcomes on a consistent basis.

To summarize, given the scale, complexity, and operational significance of their work, IDT provides an excellent case to learn from for examining inefficiencies and improvement opportunities in the current DoD organizational context. These inefficiencies are identical to the ones DoD's DE transformation policy goals aim to alleviate (Department of Defense, 2023). Perhaps more importantly, IDT handles over eighty post-production design change efforts per year for various Navy assets. Thus, while post-production design change projects are not their sole responsibility, these projects have a significant impact on maintaining operational readiness within budget limits and make up a large portion of IDT's workload. In short, IDT presents a representative and well-suited case for investigating SE inefficiencies and DE improvement opportunities, in post-production design change processes and beyond. Next, we discuss data collection.

Both qualitative and quantitative data were collected to gain a comprehensive understanding of the process. The data collection primarily involved site observations and semi-structured interviews with IDT leadership. This data was synthesized to construct two Unified Modeling Language (UML) Sequence Diagrams (Al-Fedaghi, 2021; Weilkiens, 2011) representing the design change process, capturing the tasks performed by IDT, the interactions among these tasks (energy, material, and information flows), and the corresponding stakeholders responsible for executing each of the tasks. The use of UML sequence diagrams enabled the research team to develop a sound understanding of the generic process sequence without altering the natural behaviors of the involved participants; furthermore, it was utilized to lead targeted discussions with IDT leadership for verification purposes, along with serving as a guide for quantitative data collection.

The quantitative data collection focused on expert beliefs about task durations within the identified process sequence. There was a need for relying on expert beliefs for data collection because neither IDT nor its parent organization has been collecting data on the specific task durations and deviations. Thus, there were no objective centralized data resources to draw the task duration data from. Expert beliefs about task durations were captured using triangular probability distributions, using minimum, mode, and maximum values to fit distribution characteristics. Triangular distributions were preferred due to their intuitive structure and ability to facilitate easier communication with diverse stakeholders about uncertainty and variability, and also because they are used commonly in the engineering design literature (Topcu & Mesmer, 2018). In our data collection, task duration distributions were chosen instead of other SE relevant metrics due to their: (i) critical role to determine project schedule as a limited resource in complex engineered system development (Topcu & Szajnfarber, 2025) and (ii) strong correlation with cost in engineering activities (Government Accountability Office, 2011, 2017, 2018). This concluded our data collection, and we proceeded to the qualitative analysis of the process.

### 3.2. Qualitative Analysis of the Process and Characterization of Inefficiency Modes

The next phase of the study involved a qualitative analysis of the collected data to better understand its structure and the challenges within the preliminary design phase of post-production design change tasks. As discussed above, organizations of the data into sequence diagrams revealed the interdependent nature of the design workflow, where the effectiveness of each subsequent step relies heavily on the successful completion of the preceding ones. Through this synthesis of process modeling and iterative exchange of stakeholder insights with IDT leadership, these archetypes represent patterns that consistently contribute to process delays, rework cycles, and, by extension, cost escalations. Characterizing the process in this way allowed for a structured understanding of *where, how, and why* process inefficiencies emerge. This is a convenient property of qualitative analysis because while quantitative analysis is effective in identifying the extent of process delays, it is often insufficient to uncover the underlying causes (Szajnfarber & Gralla, 2017; Topcu et al., 2021). Hence, the qualitative analysis helped to examine the mechanisms generating the inefficiencies, laying the groundwork for targeted improvement opportunities in the later phases of this study.

## 3.3. Quantitative Analysis of Schedule Deviations in the "as-is" State through a Simulation

While a qualitative analysis is valuable for identifying inefficiencies, it does not address the quantitative questions, such as how much and where digital transformation could provide the most benefit. To explore this, we adopted a quantitative approach to evaluate the process in its current state. We combined the task sequences and corresponding expert beliefs regarding the stochastic time distribution of each task to create a discrete event simulation model. Additionally, to obtain a more refined view of the process and friction points, tasks were further categorized into five categories. These categories were selected in part because they are important aspects of IDT workflows, but also because of the overwhelming evidence in the engineering design and SE literature regarding their critical role in the system lifecycle. Task categories used in this study are described below:

- *Eliciting Requirements:* Eliciting requirements involves the systematic identification, clarification, and documentation of stakeholder needs, goals, and constraints; and their postprocessing through a SE effort to establish the technical description of the problem to be solved. Once validated with stakeholders, it serves as a foundation for further design and development efforts (Mohedas et al., 2014; Wu et al., 2016; Z. Zhang, 2007).
- *Information Exchange:* Information exchange refers to the communication, transfer, and validation of data, knowledge, and decisions across project stakeholders and subject matter experts to ensure coherence throughout the system life cycle (Mesmer-Magnus et al., 2011; Rachuri et al., 2008; Robinson, 2012; Wildman et al., 2012).
- *System Level Modeling & Analysis:* This category entails the development and examination of integrated representations of the system of interest that characterize both the collective behavior of its elements and the dynamics of their interactions. These representations are expressed at a level of abstraction that reflects the system as a whole (Buede & Miller, 2024; Erbas et al., 2007; Goswami, 1997; Muller, 2012).
- *Disciplinary Modeling & Analysis:* This category focuses on the creation and evaluation of domain-specific representations (e.g., mechanical, electrical, software, chemical, electronics, etc.) to optimize subsystem design and to ensure compatibility and integration with the broader system (Buede & Miller, 2024; Shenhar, 2002; Wu et al., 2016).

- *Review Meetings:* Review meetings are formal sessions within the stage-gate structure of SE, in which stakeholders collectively assess system progress, evaluate maturity, validate requirements compliance, identify risks, and confirm that development activities remain aligned with project objectives and lifecycle milestones (D'Astous et al., 2004; Fernando et al., 2013; Liu & Yetton, 2007).

We aggregated the sequence structures, beliefs regarding task durations, and labeled the task categories in the process in a discrete-event simulation. Here, a single run of the simulation generated the total time required to complete the full process, measured in days, as well as category-level durations. To account for the inherent variability and uncertainty in the process, we conducted a Monte Carlo analysis of this simulation model. In conclusion, the simulation enabled the investigation of the extent and distribution of delays (by measuring them against the expected values), along with in which task categories they occur, and how different elements of the process contributed to overall schedule adherence.

## 3.4. Structuring of the Hypothetical "to-be" Version of the Process Post-DE Transformation and its Comparison against the Current State

Next, the simulation model that represents the as-is state of the process was compared against a hypothetical representation of a digitally transformed version of the process (i.e., the "to-be" state). To achieve this, we utilized the methodology of using a representative model to conduct simulated experiments (Panchal & Szajnfarber, 2017; Szajnfarber et al., 2020); an increasingly popular method in the SE community that helps structure simulated experiments in a cost and time-efficient manner. We generated the to-be version of the process post-DE transformation based on the DoD's DE policy goals (Office of the Deputy Assistant Secretary of Defense for Systems Engineering, 2018) by adjusting the current process's structure to match the DoD's vision. This to-be case scenario was then executed within the same Monte Carlo simulation setting as discussed in Section 3.3. Next, a comparative analysis was carried out between the as-is state and the to-be digitally transformed process. The objective of this comparison was to document where and by how much the digitally transformed process could yield a return on investment on the project schedule.

Below, we discuss the four main assumptions made for creating this to-be case of DE transformation, along with their justification, and how these assumptions were implemented to transform the baseline simulation model. The detailed implementation specifications are described in Section 4.3.

*3.5.1. Assumption 1- Up-to-date Digital Configurations within an ASOT:* Post DE transformation, there will be a comprehensive and up-to-date digital representation of the system of interest (SOI), in this case, the Navy vessel, as well as the relevant processes, systems, and operational context in which the SOI functions. The digital models are assumed to reside within an ASOT, which would play a critical role in managing and maintaining digital artifacts across the system lifecycle.

*Justification:* This assumption aligns with the first goal of the DoD's DE Strategy (Office of the Deputy Assistant Secretary of Defense for Systems Engineering, 2018), which advocates for the transformation of traditional engineering practices toward the digital and formalized creation, curation, sharing, integration, and use of models across lifecycle phases, disciplinary teams, and organizational boundaries (Huang et al., 2020b). The assumption of having a functional ASOT also aligns with the second goal of DoD's DE Strategy (Office of the Deputy Assistant Secretary of Defense for Systems Engineering, 2018), which positions the ASOT as the centralized repository and access point for standardized models, data, and supporting documentation. Moreover, the ASOT enables version control, traceability, and configuration management of system representations as they evolve over time (Allison et al., 2023). It ensures that all stakeholders across disciplines and lifecycle phases are working with the most current and validated versions of the system model, while preserving access to historical configurations and decisions (Department of Defense, 2023).

*How it was implemented:* The existence of an ASOT with lifecycle traceability could significantly reduce the need for tasks related to manual data collection and analytical model updates by the IDT. Currently, such a capability does not exist for IDT, and executing these tasks costs IDT substantial time and resources. With an ASOT in place, a validated and current digital representation of the ship and relevant systems would already be accessible, allowing teams to work from the ASOT without repeating earlier efforts.

*3.5.2 Assumption 2 - Digital Data Sharing through the ASOT:* The ASOT will enable secure, up-to-date digital data sharing among authorized stakeholders, ensuring access to up-to-date and consistent models across the system lifecycle and organizational boundaries.

*Justification*: This assumption is aligned with the second goal of DoD's DE Strategy, which advocates for the integration of interrelated digital models into a unified ecosystem (Office of the Deputy Assistant Secretary of Defense for Systems Engineering, 2018). By virtue of this assumption, when changes and information updates occur on one end of the design team or the Navy ecosystem, collaborators should be able to view and respond to updates simultaneously, enhancing coordination and reducing the volatility often associated with evolving requirements (Zimmerman et al., 2019). This shared digital environment would mitigate one of the key sources of inefficiency for IDT: the lack of timely visibility into requirement changes. This capability will also help mitigate the time losses caused by manual information handoff practices in existing workflows. In the current practice of IDT, information sharing is conducted manually, which consumes significant time for designers, especially when the disciplinary modeling and analysis tasks are concurrently executed by multiple stakeholder groups. This manual information exchange also contributes to bias and uncertainty in information-sharing tasks.

*How it was implemented:* We assumed that the existence of digital data sharing within the ASOT would reduce the time-consuming manual handoffs between disciplinary modeling and analysis tasks. That being said, we also assume that the DE transformation efforts alone will not change the time spent in conducting the disciplinary analysis. It will only reduce the time (both in terms of magnitude and variability) to share information between disciplinary analysis tasks, which often burden designers with redundant reporting and communication tasks in the status quo. The availability and access to digital information are also assumed to enable faster, less temporally variable, and more holistic collection of requirements than the status quo. Additionally, the digital data sharing capability within the ASOT could also enable digital reviews. As a result, the review meetings would become more flexible and efficient, as digital outputs allow participants to review materials asynchronously and ahead of time, reducing the need for lengthy discussions during the meeting. Since digital reviews do not require physical coordination or travel, they would also become easier to conduct across different time zones and calendars. Thus, we also assumed that this capability would lead to shorter review meetings with less and variations.

*3.5.3 Assumption 3 – Analysis in a Digitalized Environment:* The digital transformation will provide informed analysis capabilities in a digitalized and connected environment.

*Justification:* This assumption aligns with the second and third goals of DoD's DE Strategy (Office of the Deputy Assistant Secretary of Defense for Systems Engineering, 2018) which aims to ensure up-to-date data sharing, establishing an end-to-end digital enterprise. This capability is expected to enhance the efficiency of system-level modeling tasks, especially those whose performance depends on the availability of up-to-date spatial, structural, and operational data, which are subject to numerous constraints. Here, the core assumption is that the DE transformation would enable more concurrent and proactive consideration of system characteristics that may not be identified ahead of time due to their complex and interdisciplinary nature in the current capacity. This would ensure conformance to the constraints, operational realities of the ship and mission, as well as requirements of the system being installed, reducing manual effort, designer bias, and variability in early decision-making.

*How it was implemented:* The availability of advanced analysis and computation capabilities is assumed to lead to more proactive consideration and holistic conformation to the numerous constraints from the ship's topside and requirements of the new system being installed onboard. Thus, this assumption was implemented in our analysis such that feasible design alternatives are identified faster, in a manner that they are more likely to be compliant with all relevant mission requirements. It was assumed that this would reduce the probability of design rework, which is a significant gain as these issues are often unnotified until a design review.

3.5.4 *Assumption 4 - Enterprise-wide Adoption of DE:* The DE transformation will be adopted across the entire Navy Enterprise, including all relevant stakeholders, systems, and processes.

*Justification:* This assumption aligns with two of DoD's DE Strategy goals (Office of the Deputy Assistant Secretary of Defense for Systems Engineering, 2018). The fourth goal aims to transform the current IT infrastructures and environment into an integrated DE infrastructure and environment, and the fifth goal advocates transforming the culture and workforce in adopting and supporting DE across the lifecycle, which in turn would enable model-centric collaboration, common data practices, and change-management behaviors among relevant stakeholders. As a result, with an organization-wide adaptation of DE, IDT would be able to access lifecycle mission, ship, and system information from a centralized and validated source, eliminating the delays caused by staggered document delivery from multiple stakeholders. This would ensure that requirements are more complete and consistent, reducing the need for reliance on major design assumptions that are necessary to be made as IDT waits for critical mission information (e.g., EMI requirements of the new onboard sensor).

*How it was implemented:* This assumption was implemented in our study based on the premise that ASOT would enable real-time information exchange among collaborations and improved configuration control, leading to prompt and transparent communication of information updates across organizational boundaries. As a result, requirement elicitation is assumed to become faster, less variable, and more accurate. Additionally, operational realities that are currently left outside the purview of IDT, such as maintenance schedules, crew routines, and below-deck constraints, could now be rendered digitally accessible for IDT. Moreover, it would provide collaborators continuous access to IDT's ongoing analyses, and IDT would remain updated with any changes in requirements or operational data. Consequently, the time required to schedule meetings and to converge on a viable solution during review meetings can be significantly reduced. Thus, a unified digital ecosystem with full stakeholder integration and data availability from the outset would enable earlier identification of lifecycle and structural constraints, which would in turn reduce the probabilities of reiterations. Next, we discuss our findings.

## 4. Findings

We present our findings organized in three subsections. In section 4.1, we present the sequential process along with the inefficiency archetypes. In section 4.2, we discuss the findings from the quantitative analysis of the current state, pinpointing where in the process schedule delays occur, along with the magnitude and variation of these delays. Finally, in section 4.3, we present the findings regarding the return on investment of the DE transformation for schedule gains, by comparing the hypothetical to-be case of the process against its current state.

### 4.1 Why do post-production design change projects deviate from their planned schedule?

As described in Section 3, we organized our data regarding IDT's preliminary design phase of a generic post-production design process into a sequence diagram. This sequential flow was evaluated qualitatively and discussed with IDT leadership for an in-depth understanding of bottlenecks, which were labeled as *inefficiency archetypes*. We elaborate each of these inefficiency mechanisms below, along with the respective portions of the design sequence, which we split into two for ease of presentation. Please note that, also for ease of traceability, inefficiency archetypes are presented in the order they appear in the process.

We start our presentation with Figure 2, which captures the requirements elicitation and verification process that initiates each post-production design change effort. Here, following UML notation, Figure 2 should be read from top to bottom, with time progressing downward along the vertical axis that represents the sequence of activities over time. Each lifeline (i.e., vertical line) corresponds to a stakeholder, such as IDT, various design teams, or external Navy organizations IDT collaborates with. Each box on a lifeline represents a specific action, task, or decision point performed by that stakeholder. The horizontal arrows between lifelines indicate interactions, which represent transfer of energy, material, or information.

Figure 2 starts out with the official project kick-off decision, where IDT gathers the necessary stakeholders and executes an array of tasks to obtain a verified list of requirements against which the post-production design change activity will be executed, and ends with the verified topside requirements. This involves communicating with the necessary stakeholders, along with a sequence of activities and interactions (which we will elaborate below). The requirements verification process starts with the initialization of the technical kickoff. Following this, a LIDAR scan is conducted to capture the current state of the vessel, resulting in updated data in a 3D point cloud. This data is used to create or update the 3D model of the system, capturing the current configuration of the vessel. After that, a comprehensive set of mission, system, and ship requirements is gathered and/or elicited from various stakeholders. Next, IDT creates the CAD model of the system to be installed on the vessel. Next, the tailored topside requirements are generated by IDT to align with previously collected requirements. Finally, the tailored requirements were reviewed and verified through a stakeholder review process to ensure alignment and feasibility before proceeding further. The output of this phase is a verified list of topside requirements that will guide the execution of post-production design changes.

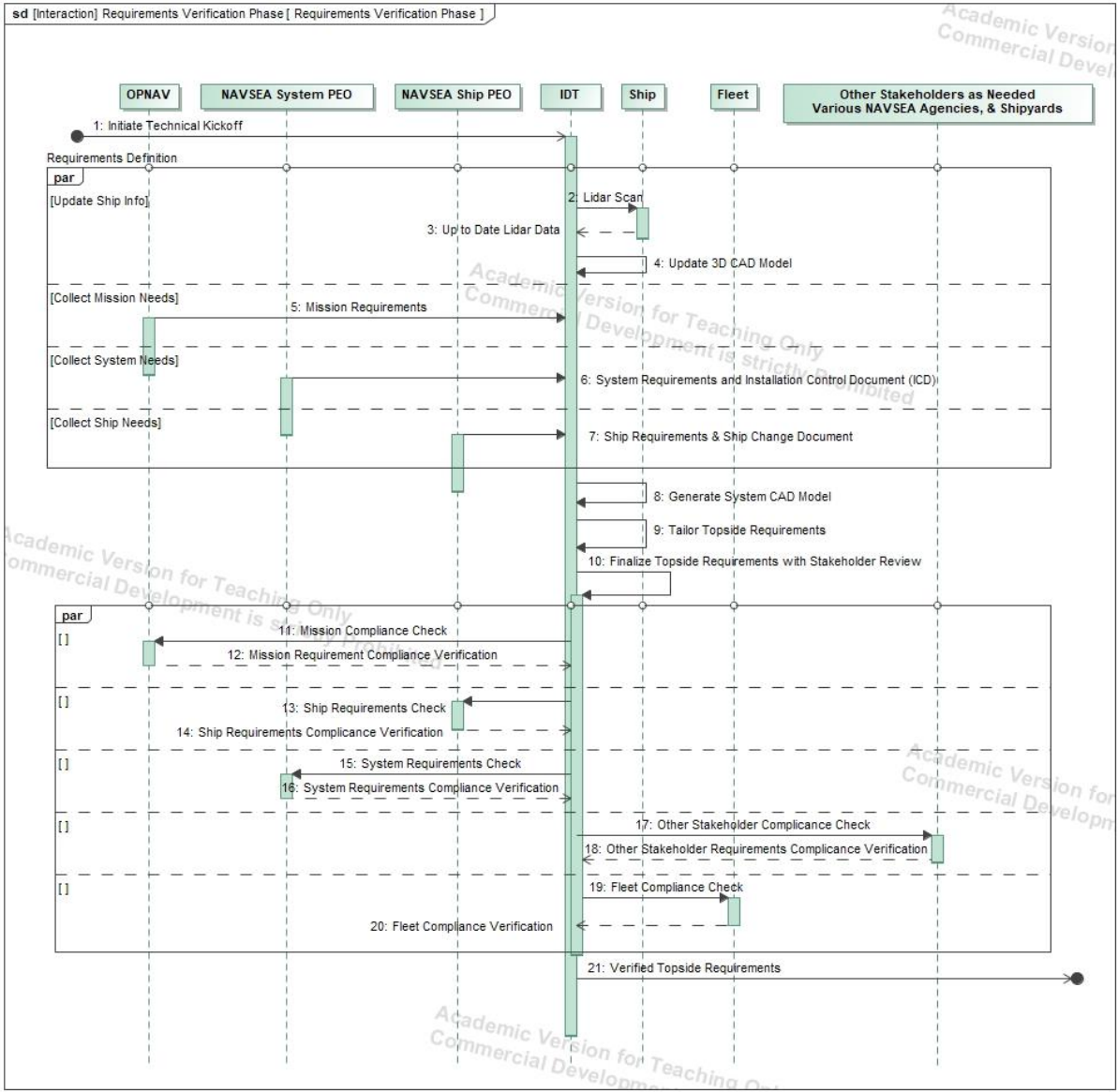

**Figure 2.** Sequence Diagram of the Requirements Verification Phase.

*4.1.1 Inefficiency Archetype 1: Lack of Up-to-Date Ship Configuration Models.*

All post-production design change projects undertaken by IDT involve existing vessels that are already in service and belong to the Navy (i.e., legacy systems). Although these ships are originally built to specific standards, the change authority lies with the ship's Captain after deployment. This decentralized authority creates significant configuration management issues, as technical baselines stored in the Naval Sea Systems Command (NAVSEA) repository often become outdated. While some of these changes may be minor, others can be substantial. As a result, it is essential for IDT to accurately determine the vessel's current configuration. To address this, IDT conducts a LIDAR scan of the vessel to capture its physical state and 3D Geometry. This scan data is then converted into a 3D point cloud, which is used to either update existing CAD models or create entirely new ones if none are available in the repository. These activities correspond to Elements #2–#4 as illustrated in Figure 2.

This inefficiency has implications for both IDT and the Navy. First, the ship has to be docked. Second, IDT personnel have to travel to the ship's location to perform the LIDAR scan. This effort can be time-consuming depending on the ship's size and complexity. Finally, the collected data must be processed and integrated into an updated CAD model of the ship. Due to the sequential nature of these steps, IDT typically tries to carry them out in parallel with external stakeholder requirements gathering (Elements #5–#7). Nonetheless, the process remains a consistent time and budget commitment that requires specialized personnel and travel.

*4.1.2 Inefficiency Archetype 2: Volatility of Mission, Platform, and System Requirements.*
In parallel with the configuration updating tasks discussed above, IDT needs information regarding the desired mission, ship, and the system (to be installed on the ship). These three pieces of information are critical for IDT to proceed. These are often provided by three (and sometimes more) different stakeholders and are captured with Elements #5-#7 in Figure 2.

This causes two issues. First, the required documents often do not arrive synchronously, and there may be delays. Second, the requirements may be volatile. Here, volatility refers to incomplete or missing information. Furthermore, in some cases, these requirements may seem complete, yet require updates and changes later in the process that are often not communicated ahead of time. Volatile (also known as moving) requirements, particularly early on in design projects, are documented to be a leading cause of delays, rework, and ultimately cost overruns (Jayatilleke & Lai, 2018; Morkos et al., 2012; Peña & Valerdi, 2015). IDT's case is no exception.

This requires IDT to adapt. IDT seeks these missing or uncertain pieces of information through external communication channels and defers further activities until they are verified. Alternatively, IDT needs to make assumptions to proceed, sometimes by consulting IDT personnel who are often in their secondary careers (with experience as Warfighters on board these vessels) who may take an educated best guess at the missing operational information. These are often useful for Ship and Mission requirements. However, if the missing information pertains to the new system to be installed – which is a frequent case, particularly if the system involves new development or emerging technologies, delays may be significant. In such cases, interface requirements may not be fully stabilized, which is a frequently observed challenge by IDT.

Thus, this inefficiency archetype impacts IDT, and by extension, the broader Navy and DoD, through the additional time and resources required to correct these activities. If the design assumptions made during this stage are inaccurate, they may go unnoticed until later in the process (e.g., a review meeting), potentially resulting in significant rework (this will be discussed further with Figure 3). Likewise, if the information provided is incomplete, such as missing critical mission requirements (e.g., target maneuverability) or system needs (e.g., stabilization requirements), it can necessitate a revision of the subsequent design process.

Once the necessary requirements are clarified, IDT aggregates them (Elements #8-#10) and organizes a stakeholder review meeting to verify the requirements (Elements #11-#20). These meetings also help to establish a shared understanding of the design objectives. In this context, the verified topside requirements serve as the starting point for the next phase of the process, outlined in Figure 3. This phase involves exploring feasible courses of action (COAs). In short, COAs describe the candidate design solutions for implementing the explored changes. Figure 3 concludes when the sponsor selects a preferred COA from the list of feasible COAs identified by the IDT. IDT then moves forward into detailed design of the final arrangement, which is out of scope of this study.

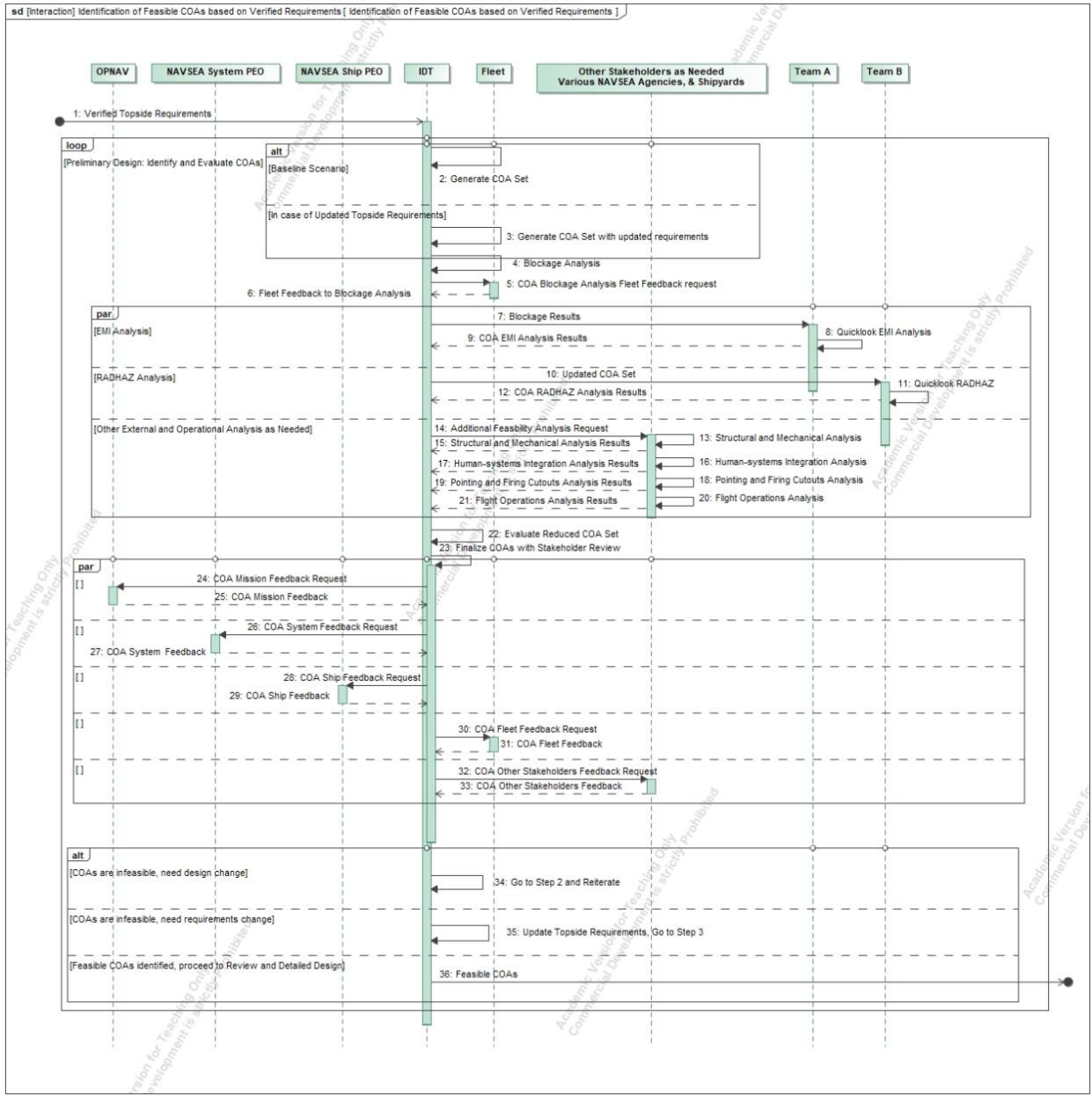

**Figure 3.** Sequence Diagram for the Identification of Feasible COAs based on Verified Requirements.

*4.1.3. Inefficiency Archetype 3: Manual Handoffs between Sequentially Dependent Disciplinary Analytical Models.*

The verified topside requirements and updated CAD models are used later in the process to identify a set of feasible COAs. This begins with 3D CAD model development of the system to be installed, guided by the system requirements and the ICD, as shown by Element #8 of Figure 2. It is followed by the identification of potential installation locations on the ship, taking the physical constraints into account (Element #2 in Figure 3). The core problem here is that the entire screening process is executed by prioritizing feasibility – by searching for where and how the new system can be placed, rather than prioritizing how it could best assist the Navy in executing its mission.

There are three issues associated with this inefficiency mode. First, the identification of initial COAs is strictly manual. This is in part because of the high number of constraints on the topside and partially due to demands from the system to be installed (e.g., physical dimensions). Given its manual nature, this introduces a significant amount of variability and bias (such as bandwagon, availability, ownership, hindsight) into the initial COA definition process based on the lead designer's knowledge and experience (Tversky and Kahneman 1974; Fillingim et al. 2022).

Second, the subsequent disciplinary analyses, as captured in Elements #4 through #20 of Figure 3, such as blockage analysis, electromagnetic interference (EMI) analysis, Radiation Hazards (RADHAZ) analysis, structural and mechanical analysis, and so on, are typically conducted in sequence. In current practice, IDT runs some of the disciplinary analysis activities in parallel, while others (for example, blockage analysis and EMI analysis) are run sequentially. Additionally, these analyses are often carried out by subject matter expert teams outside of IDT (labeled as Team A and Team B in Figure 3 to maintain anonymity), with data exchanged manually between collaborators. This creates inefficiencies, as COAs that might be ruled out as infeasible by one disciplinary analysis, for example, due to RADHAZ concerns, may still be evaluated by another team performing structural analysis assuming it is still feasible. This results in wasted effort and additional workload. As a result, while there is an opportunity to reduce unnecessary work by better integrating or synthesizing the analysis methods; the current separation of tasks is rooted in domain expertise and legacy workflows, and it contributes to longer timelines and frequent handoffs between designers. This is concerning, particularly given that these handoffs often require additional documentation and communication, which are widely regarded as time-consuming and burdensome by the design teams.

Finally, if a COA is found to be infeasible at the end of this process, it is dropped, and the remaining COAs are analyzed in depth, from the lens of each individual discipline. However, this judgement is not synthesized into an aggregate measure of goodness or value in terms of its relative future contribution to the mission effectiveness of the vessel it will be placed on (Collopy &

Hollingsworth, 2011). This, in return, yields a COA set that will be delivered to stakeholders, where decision-makers are indifferent between the alternatives. In other words, there is no "best" or "optimal" solution as a result of this process (Hazelrigg, 1996), only a set of feasible paths that can be pursued further. Judgement is passed on to the managerial decision-makers who lack the disciplinary knowledge.

*4.1.4. Inefficiency Archetype 4: Lack of System Lifecycle ASOT and Ship Operation Models.*
After a set of feasible COAs is obtained, it is brought forward for discussion in a second stakeholder review meeting (Elements #24–#33 in Figure 3), which typically involves a broader group of participants, including warfighters and shipyard representatives. However, this stage introduces two inherent challenges.

First, although key technical stakeholders from the Navy, such as NAVSEA and Office of the Chief of Naval Operations (OPNAV), are engaged early in the process to help define and constrain the COA set, their input often lacks details related to the ship's day-to-day operations, including maintenance schedules and crew routines. To address this gap, IDT relies on its internal personnel, many of whom are retired warfighters now serving in second careers. As Navy vessels operate in a harsh environment, issues like corrosion and structural wear and tear are common occurrences that require routine maintenance activities (e.g., weekly cleaning of the deck). However, these operational realities are not considered during COA identification until the Structural Analysis phases, and even then, they are not always modeled with sufficient rigor. As a result, some COAs that may be initially identified as technically feasible based on performance criteria like blockage or system compatibility may later be discarded due to operational concerns.

Second, design information that is placed outside the organizational control zone of IDT, such as below-deck structural information, is not immediately available to IDT and is not prioritized during exploration of COAs – that is, unless IDT is immediately aware of an important design challenge. This sometimes leads to "unknown unknown" design drivers that IDT may not be aware of, such as an important stakeholder to consult with. Fundamentally, this is a data fusion issue given that this information is available at some Navy-affiliated organization, yet it is not shared with IDT proactively. This leads to situations where some COAs that may appear as attractive options during COA exploration may necessitate significant installation and manufacturing challenges later on during detailed design and implementation. These are sometimes identified during Elements #24-#33 in Figure 3 through interactions with Shipyards and other Navy agencies, but that is not always the case. Often, both of these issues are not discovered until much later in the process, given their position in the sequential chain. And when they are discovered, they lead to significant rework and overshooting of schedule estimations.

We proceed into findings from the simulation study that aimed to identify the location of, and to quantify the extent of, these variations.

**4.2 Where in the process do schedule slips occur, and how substantial are these deviations?**

We analyzed the current, as-is state of the process by following the settings described in Section 3.3. More specifically, we conducted a Monte Carlo simulation with 10,000 iterations to ensure statistical stability and convergence of the results. Several key findings emerged regarding the temporal performance of the preliminary phase of the post-production design-change process as it exists today. We start the presentation of findings with Figure 4, which visualizes the distribution of project schedules measured in days, along with the spread of the task categories as boxplots. In Figure 4, the x-axis describes the number of days to complete the project or each of the task categories, and the y-axis is dedicated to labels, highlighted with colors plotted in decreasing order of their impact on the overall timeline. Recalling from Section 3, these task categories are: Eliciting Requirements, Information Exchange, Disciplinary Modeling and Analysis, System Level Modeling and Analysis, and Review Meetings.

Figure 4 reveals that the total project completion time exhibits a right-skewed distribution, with a median duration of 196.1 days and an average of 205.2 days. The project completion time has a standard deviation of 48.2 days. The higher mean relative to the median reflects positive skewness, also highlighted by a large number of outliers concentrated on the right tail. Furthermore, the spread of the outliers is noteworthy as they vary drastically. The outliers extend from 103.2 days at the lower bound to 935.9 days at the upper bound; suggesting that in certain scenarios of the current state, project timelines can experience significant delays.

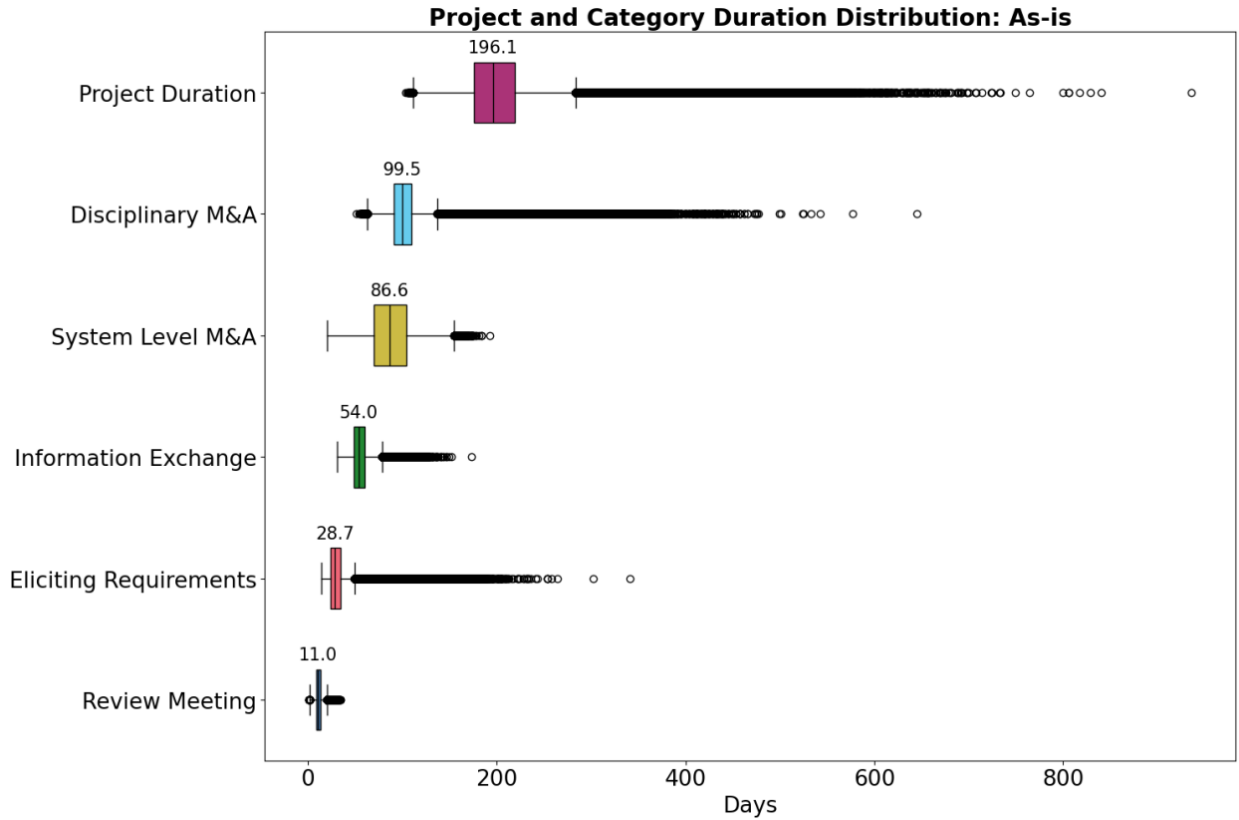

**Figure 4.** Summary of Process Completion Time and Task Category-wise Temporal Performance.

An investigation of task categories provides a refined understanding of the sources of these delays and variability. Among these, Disciplinary Modeling and Analysis contributes the most to the total project duration and variability, with a median execution time of 99.5 days and an average of 107.3 days. It also shows the highest variability, with a standard deviation of 31.6 days and the widest spread among all the categories, ranging from 51.0 to 645.1 days. The boxplot of this category reveals a right-skew and a dense concentration of outliers, similar to the boxplot of total project duration. The System Level Modeling and Analysis category is the second largest contributor to the project duration, with a median of 86.6 days and an average of 87.2 days. This task category also demonstrates high variability with a standard deviation of 23.8 days. However, its range, from 20.3 to 193.0 days, and the visible right-skewed tail in the boxplot indicate moderate spread and fewer outliers as compared to Disciplinary Modeling and Analysis, Information Exchange, and Eliciting requirements. The third largest contributor to the project duration is Information Exchange, which shows a median of 54.0 days and an average of 55.2 days. As demonstrated by the shorter box length, this category has a comparatively lower standard deviation of 9.6 days.

However, this category has a wide range from 31.4 to 172.9 days, as well as a notable number of right-tail outliers, which indicates that occasional projects may take considerably longer than typical. Next, Eliciting Requirements has a median of 28.7 days, an average of 32.2 days, and a standard deviation of 15.2 days, spanning 14.3 to 341.2 days. Although the majority of cases fall within a relatively tight spread around the median, the boxplot reveals a long right tail with a high number of outliers. Finally, Review Meetings are the shortest and most consistent category, with a median of 11.0 days, an average of 11.1 days, and a standard deviation of only 3.4 days. The narrow box and minimal outliers in this category confirm that review activities are relatively better structured and more predictable compared to other categories. However, we must note that this could be an outcome of our case selection, as in the preliminary design phase in post-production design change efforts, there is significantly less material to review compared to later stages of development, such as detailed design.

The significant variability observed in the task duration distributions suggests that the actual time it takes to complete these tasks can vary a lot from their expected value, and also from one project to another. Several factors help explain this unevenness. First, the wide range between the minimum and maximum values defined in the triangular distributions for many individual tasks works as a significant contributing factor behind the observed high variability. This broad spread allows for a greater range of possible outcomes during simulation, resulting in widely varying task durations. For example, within the Disciplinary Modeling and Analysis category, tasks such as Generate System CAD Model, Structural and Mechanical Analysis, and Flight Operations Analysis each have ranges of 30 days. The effects of such ranges are further amplified by reiterative loops, which extend task category durations unpredictably. Similarly, within System-Level Modeling and Analysis, the Update 3D CAD Model stands out as the largest contributor to variability, with a range of 113 days. By contrast, Review Meetings are bounded by a much narrower range of just 14 days, which explains why they show the least variability when compared to modeling and analysis tasks.

Another potential reason is that the skewness of the triangular distributions of the individual tasks amplifies the variability, especially when the mode lies closer to the minimum than to the maximum. This asymmetry increases the likelihood of generating values near the longer tail, contributing to variability and potential outliers. For example, in the Flight Operations Analysis task (Element #20 in Figure 3), where the difference between the mode and the minimum is only 7 days, compared to 21 days between the mode and the maximum, resulting in a right-skewed distribution that increases the probability of unusually long durations. A similar pattern is evident in the Update CAD Model task (Element #4 in Figure 2); the difference between the mode and the minimum is 53 days, while the difference between the mode and the maximum is 60 days, again favoring high variability. By contrast, the Review Meeting tasks (Elements #11-#20 in Figure 2) in the Requirements Verification Phase are much less skewed, with only a 2-day gap between the mode and the minimum and a 4-day gap to the maximum. This smaller imbalance reduces the probability of unusually long durations, making Review Meeting less variable than modeling and analysis task categories.

We further discuss these findings in Section 5, along with how our case selection and data representation could be influencing the findings. Next, we proceed to present our results from the ex-ante analysis of how DE transformation could help reshape this process.

**4.3 If the organization were digitally transformed as envisioned, where, how, and to what extent would the return on investment of DE be realized on project schedule?** This subsection presents findings on the potential return on investment of DE transformation efforts by comparing a hypothetical DE-transformed version of the IDT process against its current state. As described in Section 3.4, we implemented the DE transformation assumptions to create a future case of the enterprise post-DE transformation by modifying the baseline case. Table 1 recalls the assumptions and summarizes how each of the assumptions was implemented in the current state.

Table 1. Mapping Assumptions to DoD DE Policy Goals and Case Implementation.

| Assumptions | Corresponding DoD DE Policy Goal | How it was Implemented in this Study |
|---|---|---|
| *1) Up-to-date Digital Configurations within an ASOT* | **Goal 1:** Formalize the development, integration, and use of models to inform enterprise and program decision making. | • Tasks including docking vessels, performing LIDAR scans, and converting data into CAD models (Elements #2–#4 in Figure 2) would be replaced by validated models in the ASOT; thus, unnecessary. |

| | Goal 2: Provide an enduring ASOT. | • Similarly, as the CAD model of the system to installed would be delivered through the ASOT, IDT would not need to generate it. Element #8 in Figure 2 would no longer be necessary. |
|---|---|---|
| 2) *Digital Data Sharing through ASOT* | Goal 2: Provide an enduring ASOT. | • Up-to date data sharing within the ASOT would reduce the time-consuming manual handoffs between disciplinary functions. Particularly between Elements #4-#21 in Figure 3. It would also reduce the time for requirements collection (Elements #5–#6 in Figure 2).<br>• The availability and access to rich digital information would reduce the probability of requirements change later in the process from 10% to 7% (Element #35 in Figure 3). We believe this is a reasonable assumption because access to traceable rich digital repositories, along with continuous and transparent stakeholder integration would result in more comprehensive, consistent and accurate requirements elicitation.<br>• This would enable digital reviews, reducing the time of review meetings by 30% (Elements #11–#20 in Figure 2 and Elements #24–#33 in Figure 3). We believe this time reduction will result from asynchronous and proactive stakeholder reviews, made possible despite differences in geographical location and time zones. |
| 3) *Analysis in a Digitalized Environment* | Goal 2: Provide an enduring ASOT.<br>Goal 3: Incorporate technological innovation to improve the engineering practice. | • Advanced digital analysis capabilities with more elaborate consideration of design constraints would enable faster and more holistic identification of COAs (Elements #2–#3 in Figure 3), ensuring greater compliance with interdisciplinary constraints and reducing designer bias. This would reduce the probability of design rework from 5% to 3% (Element #34 in Figure 3), since issues are more likely to be detected and resolved earlier rather than during late-stage design reviews. |
| 4) *Enterprise-wide Adoption of DE* | Goal 4: Establish a supporting infrastructure and environments to perform activities, collaborate and communicate across stakeholders. | • Continuous organization wide stakeholder integration and transparent information updates would support faster and less variable requirement elicitation; and reduce the time required for review meetings (Elements #11–#20 in Figure 2 and #24–#33 in Figure 3). Additionally, this would support the reduction of the probability of design rework and requirements change (Elements #34 and #35 in Figure 3). |

| | Goal 5: Transform the culture and workforce to adopt and support DE across the lifecycle. | |

These assumptions were implemented to obtain the hypothetical "to-be" case of the process post DE transformation. Then, another Monte Carlo analysis was initiated to compare relative process gains against the current state of the process presented in Section 4.2. Figure 5 illustrates a comparison between the current state (as-is) and the to-be case post DE transformation in terms of overall project completion time, presented in boxplots.

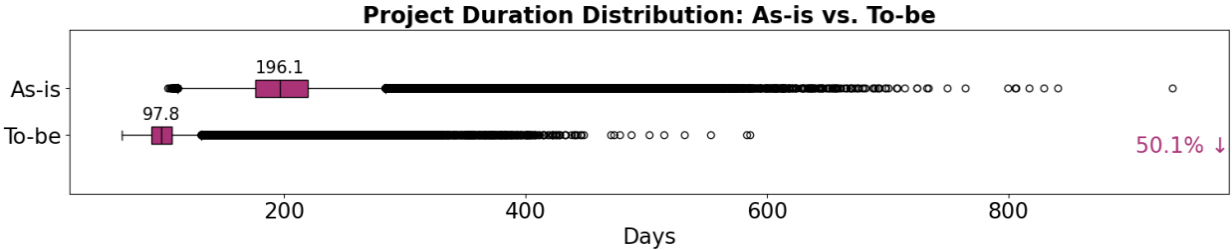

**Figure 5.** Comparison of Project Completion Time between the As-is vs. the To-be cases.

In Figure 5, the difference between the two boxes indicates a significant reduction project schedule, with the median project completion time decreasing from 196.1 days to 97.8 days, a 50.1% reduction. Furthermore, the average schedule dropped from 205.2 days to 104.2 days, reflecting a 49.2%. gain. We also observed a noticeable decrease in variability, as indicated by a more symmetric distribution as well as a tighter spread of the quartiles. The standard deviation drops from 48.2 days to 28.2 days, reflecting a reduction of 41.5%. The distribution bounds reinforce this trend. In the to-be case, the range becomes much tighter, from 65.3 days (as opposed to 103.2 days in the as-is case) at the lower end to 585.5 days compared to 935.9 days in the as-is case at the right tail. The number of outliers also decreases in the to-be case, with no outliers in the left tail. Although some variability remains, the decrease in both the upper and lower limits, standard deviation, and outliers suggests that projects are expected to finish sooner on average and are much less likely to exhibit extreme schedule overruns.

While total time saving in project completion time provides a useful overall indicator, it does not reveal where within the process schedule gains could be realized. To address this, we present Figure 6, a task category-wise comparison between the as-is case and the to-be case, where the x-axis represents the number of days, and the y-axis represents the boxplots of each task category in both states.

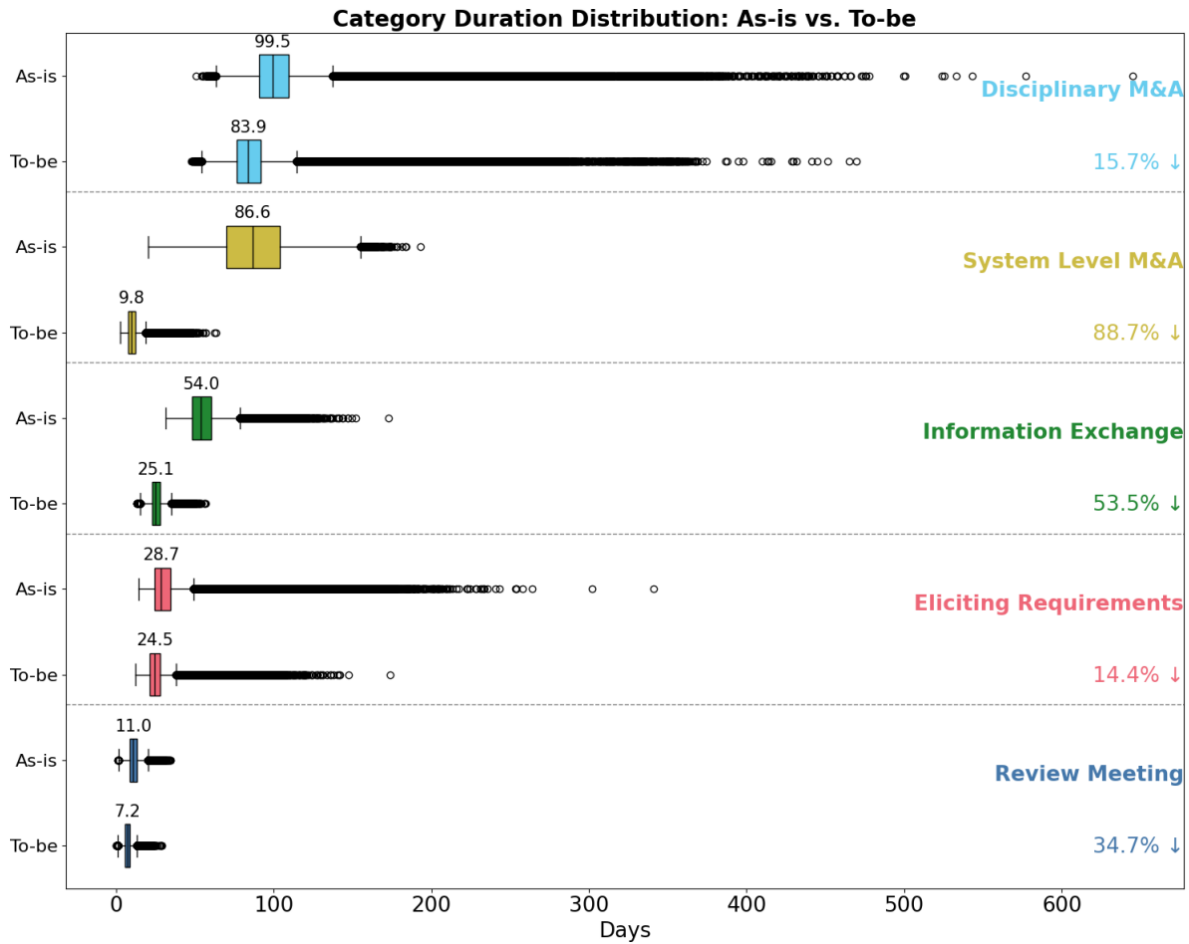

**Figure 6.** A category-wise Comparison between the As-is and To-be Cases.

Figure 6 suggests that the most significant improvement occurs in System-Level Modeling and Analysis, where the median duration decreases from 86.6 days to 9.8 days, an 88.7% reduction. The average duration similarly drops from 87.2 days to 10.5 days, and the standard deviation declines from 23.8 days to 3.9 days. This sharp decrease in task duration indicates that DE adoption could have the strongest effect on System Level Modeling and Analysis tasks, primarily because centralized access to up-to-date digital configurations and data sharing within the ASOT removes or reduces the time and variability of a large number of subtasks within this category. Among these, the largest contributor is the removal Element #4 in Figure 2, updating the 3D CAD model of the vessel, which has a mean of 62.3 days and a standard deviation of 23.1 days.

Information Exchange exhibits considerable gains, with the median falling from 54.0 days to 25.1 days, a 53.5% reduction, and the average from 55.2 to 25.3 days, a 54.2% reduction. The standard deviation drops from 9.6 to 3.9 days, and the overall range tightens significantly from 31.4–172.9 days to 13.1–56.9 days. These gains are primarily due to reduced manual handoffs between disciplinary analysis tasks. The Review Meetings task category exhibits relatively modest improvements. The median reduces from 11.0 to 7.2 days, a 34.5% reduction; and the mean drops from 11.1 to 7.3 days, a 34.2% reduction. The standard deviation drops from 3.4 to 2.3 days, and the range tightens from 1.2–34.6 days to 0.2–29.1 days. These gains are primarily enabled by the assumptions of digital review capabilities that allow asynchronous participation and pre-review of materials, reducing the need for lengthy in-person sessions.

The Disciplinary Modeling and Analysis category is one of the least improved, and remains the most time-consuming task in the to-be case as well. Nevertheless, the gains are formidable. The median duration decreases from 99.5 days to 83.9 days, a 15.6% reduction; and the average drops from 107.3 to 89.5 days, a 16.6% reduction. Variability decreases as well, with the standard deviation falling from 31.6 to 25.7 days, and the range narrowing from 51.0–645.1 days to 47.7–469.6 days. Despite these improvements, this category retains the largest spread and the most concerning right-tail outliers. This was expected, as we assumed DE transformation would not improve the time to execute disciplinary analysis tasks (except for the case of generating the system CAD model).

The least improvement was observed in the Eliciting Requirements category. In this category, the median drops from 28.7 days to 24.5 days, a 14.6% reduction; and the average drops from 32.2 days to 26.1 days, suggesting an 18.9% reduction. However, a larger improvement is observed in the standard deviation, as it decreases from 15.2 days to 8.2 days, indicating a substantial reduction in variability. The lower and upper bounds narrow from 14.3–341.2 days to 12.4–173.7 days. The improvement is enabled by digital access to more holistic information regarding the requirements and constraints. At the same time, the lesser improvement as compared to other categories indicates that only adapting DE without changing the organizational culture is not enough to realize the full potential of DE. Next, we discuss our findings.

## 5. Discussion

This study examined why post-production design change projects run late (RQ1), where delays arise and how substantial they are (RQ2), and to what extent a digitally transformed future state could improve outcomes (RQ3). Focusing on the preliminary design phase of an IDT at an anonymized Navy agency, we combined a qualitative analysis with quantitative simulation models to move from understanding the problem in the current state to estimating the potential performance gains achievable through digital transformation. Below, we discuss our findings in terms of their significance, limitations, and implications for policymakers, managers, and researchers.

### 5.1 Significance of the Findings

This study makes three core contributions. First, to the best of our knowledge, it is the first attempt to explain in detail why post-production design change projects routinely overrun their projected schedules. Our findings qualitatively formalize these deviations in four *inefficiency archetypes*: manual information handoffs, volatile requirements, lack of up-to-date configuration models, and lack of lifecycle ASOT and operation models. Although none of these archetypes are surprising, the key contributions are: (i) the rich documentation and explanation of how these inefficiencies manifest themselves, and (ii) their broader generalizability to other SE applications and organizations. Many of these inefficiency archetypes are characteristic of the sociotechnical nature of conducting post-production design change activities on large-scale complex legacy systems.

For instance, the absence of a system-lifecycle ASOT, poor configuration control, fragmented analytical models, and/or data interfaces consistently produce similar patterns of rework and queuing issues across domains. These conditions lead engineers to initiate tasks without reliable, up-to-date baselines, which in turn either delay the process significantly or allow it to proceed based on strong design assumptions that introduce additional uncertainty into the process. This results in additional downstream engineering effort, and inconsistencies between various interrelated disciplinary analyses are often detected too late into the process, which leads to the manifestation of more rework. In contrast to these technical process related mechanisms, some of the inefficiencies are rooted in local practices and organizational culture, such as the frequency of cross-team reviews, the allocation of information exchange and decision rights, and the coordination of design interdependencies across teams and stakeholders. Understanding these inefficiencies and their sociotechnical nature is essential for targeting interventions at specific points in the process where they can most effectively reduce delays and improve overall performance.

The second major contribution of this study is the documentation of *where* and *to what extent* schedule delays occur. We found that task categories contribute unevenly to the total project time, and their predictability also differs drastically. Some categories, such as System Level Modeling and Analysis, Disciplinary Modeling and Analysis, and Information Exchange, dominate the schedule. At the same time, Eliciting Requirements and Review Meetings play smaller but still consequential roles. We must note that some of this heterogeneity is expected, and the variability in relative durations in this study primarily stems from the characteristics of the process under investigation. The preliminary design phase carried out by IDT for post-production modifications is a tightly coupled system, and the engineering teams leading these efforts are often not the original designers of the vessel nor the system to be installed on the vessel. As a result, they must invest substantial time in System Level Modeling and Analysis to ensure proposed changes do not disrupt existing functions of the legacy system nor violate any of its constraints. This also makes the Disciplinary Modeling and Analysis a major contributor because COAs need to be tested for feasibility against a series of domain-specific analyses. Moreover, the involvement of multiple internal and external stakeholders and collaboration between teams make information exchange a significant contributor to total project time. By contrast, although Eliciting Requirements and Review Meetings remain critical for ensuring design integrity and stakeholder alignment, their relative time contributions are smaller compared to the Modeling and Analysis and Information Exchange categories that drive most of the project duration. This is mostly because we based our analysis on a preliminary design phase, as opposed to a detailed design or system test and evaluation, which usually involves a much larger set of design artifacts to be reviewed.

Additionally, a significant variability in both project and task completion times was observed, with a pronounced right skew and a high number of outliers. This indicates that there is a non-negligible probability of extreme delays, which disproportionately affect project planning. This finding is in line with the empirical reality of today. For instance, as we are finalizing this manuscript, the U.S. Navy's aircraft carrier fleet is facing a severe readiness crisis due to systemic, multi-year delays in O&S. Among these, perhaps the most significant one is the Nimitz-Class Nuclear Aircraft Carrier USS John C. Stennis, which is declared out of action for "five to six years" due to midlife O&S activities. Although not all of this downtime is spent during preliminary design, it has been well established in the SE community that these early-stage design decisions dictate the goodness of development outcomes, in both novel development and O&S alike. To that end, this paper sheds light on how these "extreme" delays are unfolding.

In our case, some of the observed variability stemmed from inherent uncertainty in task completion times, depending on external stakeholder availability and differences in complexity and path dependencies. The variability was further exacerbated by the manual and judgment-based nature of many IDT tasks. For example, tasks involving information exchange are highly dependent on timely input and coordination between multiple stakeholders. On top of that, natural variation in human performance in information exchange tasks makes them particularly susceptible. This task level variability often propagates upward, amplifying delays across task categories and ultimately extending total project duration.

Perhaps more importantly, the third major contribution of this study is that it offers the first quantification attempt at *where* and *how* DE transformation can generate its return on investment. It has been more than seven years since the DoD published its DE policy goals, and the full-scale implementation of these policy goals across the DoD enterprise has been lackluster. Although DE's purported benefits are promising, this slow adoption is in part because both the managers and engineers in the trenches of various DoD agencies have no substantial basis for how exactly DE could help them handle their daily work. This paper presents evidence regarding which aspect of the process DE improvements could be expected and by how much they could contribute. In our case, task categories that depend heavily on system representations and sharing information across organizational boundaries, such as System Level Modeling and Analysis, Information Exchange, and Review Meetings, are expected to realize the largest gains. These substantial gains suggest that DE's strengths -at least for this phase of the development cycle- lie in enabling integrated data environments, reducing manual handoffs, and improving the accessibility and consistency of digital models and information across teams. Conversely, relatively modest improvements (~15%) in Disciplinary Modeling and Analysis and Eliciting Requirements indicate that DE alone cannot fully overcome domain-specific or judgment-intensive tasks that necessitate SE expertise. To realize meaningful improvements in these categories, the organization must also embrace a cultural and behavioral shift in how it collaborates and shares information in the ASOT.

In conclusion, although this research focused on a single organization under conservative assumptions, the projected gains are substantial, indicating the potential for significantly greater returns on investment if DE practices are scaled enterprise-wide. We must also note that gains documented in this study are only projected to be achieved via DE transformation alone. Thus, the use of generative artificial intelligence (AI) in the form of cognitive assistants and other SE tools could provide new opportunities; which could be enabled by large multi-modal data synthesis by DE transformation promises. These AI technologies could also amplify the DE-only gains documented here. In short, this study provides strong evidence for the claim that DE could significantly help SEs navigate the golden triangle of performance, cost, and schedule (Office of the Deputy Assistant Secretary of Defense for Systems Engineering, 2018; Topcu & Szajnfarber, 2025).

### 5.2 Limitations

While this study provides insights into the potential impact of DE transformation on post-production design change processes, several limitations that may limit the broader generalizability of the findings should be acknowledged.

First, this study focuses exclusively on the project schedule and does not account for cost dynamics. While there is a strong correlation, time savings do not necessarily translate directly into cost savings because factors such as variability in labor rates, workforce seniority, and resource allocation can influence overall financial impacts. Regardless, this was a necessary research choice on our end to take the first step towards quantifying the return on investment of DE transformation.

Second, the scope and complexity of projects vary significantly; however, this study focused on the structure of the design process and collected expert beliefs regarding task distributions in aggregate. In reality, the projects carried out by IDT differ considerably in scale, scope, and complexity. For example, a minor modification on a small vessel, such as a coastal patrol boat, poses much more manageable design complexity, stakeholder coordination, and verification demands compared to a major upgrade on an aircraft carrier. These differences could further amplify the variations in findings. However, this was a necessary research choice as the Sponsor lacked a centralized data collection source, and it made little sense to conduct several different versions of this study for varying levels of vessel complexity. Additionally, these factors were present on both the current state of the process presented in Section 4.2 and the hypothetical future state results presented in Section 4.3; thus, these variations due to vessel specifics should not alter the main findings of this study. This claim was also supported by our debriefing discussions with the Sponsor.

Third, some of the relative duration of these task categories and their variability stems heavily from the nature of the process under investigation. The preliminary design phase for post-production changes involves a tightly coupled system, which may or may not be designed by the current IDT personnel. This leads to certain tasks, such as system-level analysis, taking a significant amount of time compared to the other categories, to avoid unintended consequences. Hence, while the findings documented in this study provide valuable insights into the specific IDT context, the specific percentage gains should not be interpreted as universally prescriptive for all post-production design change processes or organizational settings. Nevertheless, while the nuances and specific projected gains are expected to vary, the general findings should hold true for most preliminary design processes, particularly those conducted during sustainment of large-scale legacy systems.

## 5.3 Implications for Managers, Policymakers, and Researchers

For *practitioners and managers*, our findings underscore the importance of developing an in-depth understanding of the underlying challenges, bottlenecks, and organizational dynamics that shape the performance of the complex SE projects. Such understanding is essential for ensuring that DE adoption targets the structural sources of inefficiency, rather than treating symptoms in isolation. Effective practice should therefore focus on addressing organization-specific issues, including weak configuration control, fragmented data infrastructures, and limited lifecycle visibility, so that DE implementation aligns with actual process needs and operational constraints. This study provides evidence that the benefits of DE adoption are neither uniformly distributed across all processes nor all tasks the organization is responsible for. Hence, understanding where DE delivers the greatest impact can help practitioners prioritize high-leverage investments and focus transformation efforts where they yield the most value, particularly given DE's high upfront costs.

For *policymakers,* especially for those who influence government policy, this study is the first of its kind that presents evidence for how DE transformation could effectively improve large-scale complex system development efforts. We contend that a clearer understanding of these benefits could help provide the necessary impetus for negotiating and securing the necessary budgetary resources to help enact the DE policy. While the DoD has made significant progress in setting expectations through guidance documents, without long-term financial commitment, programmatic support, clearly defined success metrics, and their proactive tracking, it will not be possible to realize long-lasting DE transformation. Thus, prioritizing research support and funding to generate empirical evidence for DE benefits is essential for stakeholder buy-in. A more subtle but important policy implication of this study is how cross-organization interactions and distribution of domain expertise influence program-level outcomes, along with the workload of engineers in the trenches. Although it will not be a smooth transition, DE transformation could greatly simplify this complexity, and quite possibly lead to unexpected efficiency gains. To alleviate these obstacles, we recommend investing in a fully digital infrastructure, continuous enhancement of workforce capabilities, and creating sustainable support mechanisms that enable organizations to adopt and scale DE effectively.

For *researchers,* this study lays the groundwork to move DE beyond a theoretical idea by providing measurable and refutable evidence regarding where and how it could yield its benefits, and eventually help enhance SE performance. As DE reshapes how complex engineered systems will be developed and managed, researchers should advance evidence-based strategies for implementation, policy design, and impact evaluation of DE adaptation. Priorities include developing validated metrics and running reproducible mixed-methods studies (comparative cases, simulation, etc.) to assess how DE practices like MBSE, traceability, and configuration control with the ASOT influence schedule, technical system performance, cost, and risk. Such work can help bridge the gap between DE's promise and its demonstrated impact on SE performance.

Finally, although this study did not focus on it, DE will present a tremendous opportunity to position AI for SE and SE for AI initiatives. Related to this, most of the recent advances in the AI community have been primarily due to the broad availability of data, yet the SE community greatly suffers from data availability, as well as benchmarks for evaluating the effectiveness of new methods. We contend that the community could greatly benefit from more efforts in these research gaps, which could collectively move the field forward, particularly in the development and testing of new methods.

## 6. Conclusion

Complex systems development and sustainment efforts are routinely over budget and schedule. The SE community has been increasingly looking into DE as a remedy to these acute problems. Yet, despite seven years having passed since the DoD DE Policy Goals, DE adoption has been slow in both the government and the industry, particularly because its potential return on investment is poorly understood. This study adopted a mixed-methods approach to explain why system sustainment efforts routinely suffer from schedule slips, documented how substantial these deviations could be, and provided the first-ever quantification of how DE transformation could help reduce these delays. These findings suggest that DE transformation could substantially improve schedule performance, although its benefits may be unevenly distributed. Some SE tasks are expected to realize greater gains than others, underscoring the need for differentiated implementation strategies. Together, these results can equip decision-makers with clear, data-driven guidance on where and how organizations should prioritize DE adoption to achieve the greatest impact.

Building on the findings of this research, future work can prioritize several alternative directions. One avenue is conducting a sensitivity analysis to explore different DE adoption scenarios and their potential effects. Additionally, expanding the analysis to multiple organizations and different phases of the system lifecycle would help validate the broader applicability of our results. Incorporating a comprehensive cost-benefit analysis of DE implementation could provide a more holistic understanding of its return on investment, enabling decision-makers to evaluate trade-offs between time, cost, and resource optimization; however, this would necessitate significant data collection efforts. Finally, we contend that DE transformation will provide new opportunities for both AI for SE and SE for AI research. While this study did not focus on this research thrust, we look forward to seeing new advances in this area.